# Spin wave non-reciprocity and beating in permalloy by time-resolved magneto-optical Kerr effect

Siddharth Rao, Jungbum Yoon, Jan Rhensius, Charanjit S. Bhatia, and Hyunsoo Yang[a]

*Department of Electrical and Computer Engineering and NUSNNI, National University of Singapore, 117576, Singapore*

We have studied the propagation characteristics of spin wave modes in a permalloy stripe by time-resolved magneto-optical Kerr effect techniques. We observe a beating interference pattern in the time domain under the influence of an electrical square pulse excitation at the center of the stripe. We also probe the non-reciprocal behavior of propagating spin waves with a dependence on the external magnetic field. Spatial dependence studies show that localized edge mode spin waves have a lower frequency than spin waves in the center of the stripe, due to the varying magnetization vector across the width of the stripe.

[a] E-mail: eleyang@nus.edu.sg

# 1. Introduction

Spin waves are defined as propagating disturbances in the magnetic ordering of materials such as ferromagnets, ferrimagnets and antiferromagnets via exchange or magnetostatic interactions. The characteristics of spin waves can be tuned by varying the static magnetization ($M$) and the direction of spin wave propagation ($k$). These properties have generated a great deal of interest in spin waves due to their potential applications in high-speed information processing [1-6], synchronization of spin torque oscillators [7, 8] and enhancement of the spin pumping effect [9]. Recent reports of interference-mediated modulation of spin waves have highlighted these capabilities by providing an additional tuning parameter in designing such circuits [10, 11]. Logic gates based on spin waves have been proposed and experimentally demonstrated [12, 13]. Hence, it is important to understand the magnetization dynamics on a nanosecond time scale in greater detail, as most technological applications will operate in this regime.

The experimental study of spin wave dynamics requires the use of systems with high temporal and/or spatial resolution. Electrical techniques such as spin wave spectroscopy are limited in their spatial resolution, while optical techniques such as micro-Brillouin light spectroscopy (micro-BLS) offers impressive spatial resolution (~250 nm) [14] but moderate temporal resolution. Spin wave excitation and propagation characteristics in laterally confined magnetic stripes using micro-BLS have been extensively studied in recent years [14-21]. The finite size effect due to the lateral confinement of spin waves results in interference between spin wave modes revealing a spatial dependency of the spin wave intensity [14, 16]. Time-resolved scanning Kerr microscopy (TRSKM) has a significantly higher temporal resolution (~20 ps) than BLS measurements, and this advantage has been used to explore the temporal characteristics of spin wave propagation [22-26]. Spin waves have been detected at distances up to 80 μm away from a coplanar strip (CPS) line [23] in thin permalloy films using coplanar waveguides antennas. Studies on thin permalloy (Ni$_{80}$Fe$_{20}$) films of different aspect ratios and geometries have reported the observation of mode interference of single-frequency spin waves in the *k*-space [23, 26, 27].

In this paper, we study the propagation of spin waves in thin permalloy (Ni$_{78}$Fe$_{22}$) films by TRSKM, and report the presence of a low frequency beat-like interference pattern at the center of the stripe due to two distinct spin wave frequencies. We provide a simple mathematical model to explain our observations, and show good correlation between experimental and simulated data. In addition, we observe a clear non-reciprocity in the spin wave propagation characteristics under the influence of an external magnetic field. The existence of different spin wave modes across the stripe is shown through spatial dependent measurements, and is attributed to the varying magnetization profile across the stripe.

## 2. Experimental Details and results

The pump-probe experiments are performed on samples of Si substrate/SiO$_2$ (300 nm)/Ta (2 nm)/ Ni$_{78}$Fe$_{22}$ (30 nm)/ SiO$_2$ (3 nm) sputtered at a base pressure of 2×10$^{-9}$ Torr. The films are patterned into 200 μm × 20 μm stripes by electron-beam lithography. A 9 μm-wide stripline of Ta (5 nm)/Cu (150 nm) is deposited and patterned over the center of the Py (Ni$_{78}$Fe$_{22}$) stripe by standard sputtering and photolithography processes. The Ta underlayer is added to improve the adhesion of the Py and Cu layers to the Si/SiO$_2$ substrate. The time-resolved measurements are performed in a transverse mode TRSKM configuration. The magnetization dynamics are probed by a 2 mW laser pulse of 45 ps pulse width and a wavelength of 405 nm. The pump-induced Kerr rotation is measured as a function of the delay time between the pump and probe pulses at room temperature, with a time resolution of 20 ps. The measured signal is proportional to the *y*- component of the dynamic magnetization under the optical probe spot (~1 µm). An external magnetic field ($H_b$) is applied perpendicular to the long axis of the Py stripe and the electrical pulse-generated Oersted field ($h_{pulse}$), as shown in figure 1(a). An electrical pulse of 10 V amplitude and 5 ns duration, with a rise time of 55 ps, is applied to the stripline to generate an Oersted field of $h_{pulse}$ = 17.9 Oe. As a result, spin waves of the Damon-Eshbach mode are generated and propagate along the ±y-direction [28]. Figure 1(b) shows a representative time response of a signal measured in this setup at the center of the Py stripe with a clear Gaussian spin wave packet envelope, while figure 1(c) shows the corresponding FFT spectrum due to the rising edge of the electrical pulse. The two main peaks in the FFT spectrum indicate the possibility of the existence of multiple modes at the same location, which will be discussed in further detail later.

In order to ensure that our measurement technique is correct, we perform experiments to measure the spin wave decay length, non-reciprocity factor, bias magnetic field dependence and spatial dependence of the spin waves in a magnetic stripe as discussed below. Figure 2(a) shows the characteristic dependence of the spin wave frequency on the external bias field ($H_b$) at the center of the Py stripe measured 10 µm away from the current-carrying stripline. The magnetic stripe saturates in the transverse direction at $H_b$ = ± 40 Oe. Each value of the bias field used for measurements is approached after saturation at a higher field in the same direction. The spin wave frequency dependence can be explained by fitting with the magnetostatic surface spin waves (MSSW) dispersion relation $f = \frac{\gamma_o}{2\pi}\sqrt{H(H+4\pi M_s)+\left(\frac{4\pi M_s}{2}\right)^2(1-\exp(-2kd))}$, where $H$ is the total effective field in the system, $\gamma_o$ = 17.6 MHz/Oe is the gyromagnetic ratio, $k$ = 0.5678, 0.1243 µm$^{-1}$, and the film thickness $d$ = 30 nm. Two frequencies are observed at every bias field value indicating multiple spin wave mode generation from the square pulse excitation, which contains various frequency components. Both modes are well-fitted with the MSSW dispersion relation with different *k*-vectors as has been reported before [29]. The slightly different frequencies between these modes are critical to the interference studies discussed further on. To understand the evolution of the center modes, we analyze the spatial dependence of the spin wave intensity. Measurements are performed at different distances from the current-carrying stripline by moving the laser spot along the long axis of the Py stripe. The spin wave amplitude (*A*) is extracted from a fitting of the measured signals in the time domain, as shown in the inset of figure 2(b) using a Gaussian wave packet of the form $y = A\exp(-x/\Lambda)\exp\left[-(t-t_o)^2/2\sigma^2\right]$, where $t_o$ is the temporal position of the peak of the Gaussian wave packet, and $\sigma$ is the full-width half maximum (FWHM) of the spin wave packet. The spin wave decay length $\Lambda$ = 15.44 μm is extracted from the exponential fitting with the formula $A\exp(-x/\Lambda)$ [29, 30], where *x* is the distance from the stripline in figure 2(b).

The amplitude of the MSSW signal depends on the relative directions of the magnetization (*M*) of the material and the spin wave wave-vector (*k*). A clear non-reciprocal behavior of spin waves is known to exist for opposite signs of wave-vectors (±*k*) or bias field (±$H_b$) as shown in figure 2(c) [15, 19, 30, 31], and the non-reciprocity factor (*κ*), defined as *κ* = $A(H_{b+})/A(H_{b-})$, is a function of the bias field as shown in figure 2(d) similar to electrical measurements of surface spin waves [13]. The higher temporal resolution of TRSKM as compared to BLS measurements makes it a suitable measurement technique to confirm previous findings on non-reciprocal behavior of spin wave dynamics in the time domain. The non-reciprocity in figures 2(c) and 2(d) is evaluated at a distance of 10 μm from the stripline. The non-reciprocal behavior arises as a result of the asymmetric distribution of the out-of-plane component of the excitation field ($h_z$) on either side of the stripline [15, 31]. The in-plane component of the excitation field ($h_y$) retains the same phase across the stripline, while $h_z$ undergoes a phase change of π. As a result, the spin waves excited due to in-phase $h_y$ and $h_z$ fields (on one side of the stripline) are stronger than those excited by out-of-phase $h_y$ and $h_z$ fields (on the opposite side of the stripline), resulting in the non-reciprocity of spin waves. The magnitude of κ is of the same order as previously reported [30, 31], and can be tuned by adjusting the width and thickness of the stripline which controls the excitation field ($h_{y,z}$). This phenomenon has promising potential in the field of logic devices to represent binary states of '1' and '0', as has been recently proposed [13].

The aspect ratio (10:1) of the magnetic stripe is another factor that can affect the spin wave characteristics due to spatial confinement [14-16, 32]. We study the effects of this parameter by carrying out measuring the spin wave characteristics at different distances from the left edge to the center of the stripe, as indicated by the red spot in the schematics of figures 3(a) and 3(b). The FFT spectra are shown in figures 3(a) and 3(b), which corresponds to the data measured at distances of 5 and 10 μm away from the current-carrying stripline, respectively. The measurement data, carried out at $H_b$ = 75 Oe and a 10 V electric pulse of 0.2 ns duration, are consistent with the existence of localized edge modes at the edges of the stripe [32-34]. Due to the shape anisotropy, the magnetization at the edges is pinned along the long axis of the Py stripe, while the magnetization towards the center of the stripe tends to align with the external bias field. This creates a varying magnetization profile across the width of the stripe [34], which has been used to explain that the spin wave dispersion relation changes as a function of position [25, 33, 35]. Due to the varying demagnetizing field profile and stronger exchange interaction effects at the edges [32, 34], a weaker effective field exists at the edges as compared to the center of the stripe and as a result, the frequencies of the edge mode are lower than that of the center mode.

## 3. Discussion

Spatial dependence studies reveal the existence of a frequency beat-like interference pattern at 8 μm from the stripline, as seen in the measured signal in figure 1(b). The FFT spectrum in figure 1(c) clearly shows the existence of two frequency peaks in close proximity. The frequency resolution of these measurements is ~70 MHz, which rules out the possibility of any artifact peaks generated due to the Fourier transform operation. Unlike previous spin wave mode interference studies performed in the spatial domain [14, 16, 27], the origin of the observed interference in our experiment is in the frequency domain. In other words, the spin wave modes causing the interference are separated by frequency, rather than by different $k$ values for the same frequency. Our result is also different from a recent study, in which an oscillatory spatial intensity pattern superimposed on the exponential decay is observed by the BLS technique [36]. On the other hand, in our case, the beating pattern can be attributed to the interference of the two center spin wave modes measured in figure 2(a). The two center modes having different frequencies will have a phase difference due to which they arrive at the same given location. We model the oscillatory beat pattern as interference between two Gaussian spin waves of slightly different frequencies arriving at a given location at the same time as

$$y(t) = y_o + B_1 \exp\left(-\frac{1}{2}\left(\frac{t-t_1}{\sigma_1}\right)^2\right)\sin(2\pi f_1 t) + B_2 \exp\left(-\frac{1}{2}\left(\frac{t-t_2}{\sigma_2}\right)^2\right)\sin(2\pi f_2 t), \text{ where } f_1$$

= 2.22 GHz and $f_2$ = 2.46 GHz are the respective frequencies of the Gaussian spin wave packets,

$t_1$ = 1.005 ns and $t_2$ = 4.37 ns are the temporal positions of the peaks of the respective Gaussian spin wave packets, $\sigma_1$ = 0.464 ns and $\sigma_2$ = 1.11 ns are the FWHM of the two spin wave packets, $B_1$ = -197.557 and $B_2$ = -120.281 are the respective arbitrary amplitudes of the two wave packets, and $y_o$ = 399.766 a.u. is the initial offset. This model fits well with the experimental time-domain data shown in figure 1(b), and the frequency values also show reasonably good agreement. Thus, the interference of the two center spin wave mode frequencies results in a unique mode-beating interference pattern in the time domain, different from previous mode interference studies that were identified in the *k*-space. These results can have implications on the proposed applications of spin waves in information processing devices, which utilize the amplitude modulation by the observed interferences.

In order to enhance our understanding of the beating pattern, we have performed micromagnetic simulations using the object oriented micromagnetic framework (OOMMF) [37]. A simulation cell size of $50 \times 50 \times 20$ nm$^3$ is chosen and standard material parameters for Py are used such as the saturation magnetization $M_s$ = $8.0 \times 10^5$ A/m, exchange stiffness constant $A$ = $1.3 \times 10^{-11}$ J/m, gyromagnetic ratio $\gamma$ = $2.32 \times 10^5$ m/(A·s) and the Gilbert damping constant $\alpha$ = 0.01. The micromagnetic simulations were performed at different cell sizes, namely $50 \times 50 \times 20$ nm$^3$ and $100 \times 100 \times 20$ nm$^3$, and similar results were obtained in terms of the temporal and frequency response. To simulate the Oersted field due to a current-carrying stripline at the center of the stripe, a pulsed magnetic field is applied to a volume of 9 μm $\times$ 20 μm $\times$ 30 nm as indicated by the dotted region in figure 4(a). The amplitude of the pulsed magnetic field is 17.9 Oe, as in the case of the experiment. An in-plane bias field $H_b$ = 100 Oe is applied along the –*x*-axis.

Figure 4(a) shows a representative snapshot of the spatial variation in $M_y$ at $t = 3$ ns. We analyze the spin wave dynamics at different distances along the center of the magnetic stripe, as shown in figure 4(b). The simulated waveforms suggest the existence of two regimes, namely ferromagnetic resonance (FMR) dynamics and spin wave dynamics, and their relative intensities depend on the distance from the stripline. We analyze the data on either side of the stripline, and the simulated waveforms show a gradual change from FMR dynamics (out-of-phase waveforms for $\pm k$) to spin wave dynamics (in-phase waveforms for $\pm k$) as reported before [13]. As we move further away from the stripline, the Gaussian envelope of the propagating spin waves is observed more prominently in comparison to the exponentially-decaying envelope of FMR-type dynamics. At a distance of 5.5 µm from the stripline, we observe a beating envelope in the spin wave propagation with time (for $t \sim 3$ ns). This beating envelope is also seen further away from the stripline, as shown in figure 4(b). Figure 4(c) shows a magnified view of the spin waves at a distance of 45.5 µm from the stripline, clearly displaying the beating envelope similar to the experimental data in figure 1(b). A Fourier transform reveals the presence of two main peaks at $f = 2.84$ GHz and $f = 3.08$ GHz at shown in the inset of figure 4(c), confirming the existence of two main frequency modes at the center of the stripe.

## 4. Conclusions

We have reported the observation of a beating pattern in time domain due to the interference of two spin waves with different center mode frequencies by experiments and simulations. We model the interference based on a simple mathematical model considering frequency- and phase-separated Gaussian spin waves, and it shows good correlation between the measured and fitting data. We have also demonstrated, in the time domain, the non-reciprocity effect of spin waves as a function of the applied bias field, a phenomenon that can be engineered to create spin wave-based logic circuits.

**Acknowledgements**

This work is partially supported by the National Research Foundation, Prime Minister's Office, Singapore under its Competitive Research Programme (CRP Award No. NRF-CRP 4-2008-06).

# References

[1]  Kruglyak V V, Demokritov S O and Grundler D 2010 *J. Phys. D Appl. Phys.* **43** 264001
[2]  Bandyopadhyay S and Cahay M 2009 *Nanotechnology* **20** 412001
[3]  Gorshkov A V, Otterbach J, Demler E, Fleischhauer M and Lukin M D 2010 *Phys. Rev. Lett.* **105** 060502
[4]  Kajiwara Y *et al* 2010 *Nature* **464** 262
[5]  Murphy M, Montangero S, Giovannetti V and Calarco T 2010 *Phys. Rev. A* **82** 022318
[6]  Buczek P, Ernst A and Sandratskii L M 2010 *Phys. Rev. Lett.* **105** 097205
[7]  Kaka S, Pufall M R, Rippard W H, Silva T J, Russek S E and Katine J A 2005 *Nature* **437** 389
[8]  Mancoff F B, Rizzo N D, Engel B N and Tehrani S 2005 *Nature* **437** 393
[9]  Mukherjee S S, Deorani P, Kwon J H and Yang H 2012 *Phys. Rev. B* **85** 094416
[10] Mukherjee S S, Kwon J H, Jamali M, Hayashi M and Yang H 2012 *Phys. Rev. B* **85** 224408
[11] Kwon J H, Mukherjee S S, Jamali M, Hayashi M and Yang H 2011 *Appl. Phys. Lett.* **99** 132505
[12] Schneider T, Serga A A, Leven B, Hillebrands B, Stamps R L and Kostylev M P 2008 *Appl. Phys. Lett.* **92** 022505
[13] Jamali M, Kwon J H, Seo S-M, Lee K-J and Yang H 2013 *Sci. Rep.* **3** 3160
[14] Demidov V E, Demokritov S O, Rott K, Krzysteczko P and Reiss G 2008 *Phys. Rev. B* **77** 064406
[15] Demidov V E, Kostylev M P, Rott K, Krzysteczko P, Reiss G and Demokritov S O 2009 *Appl. Phys. Lett.* **95** 112509
[16] Demidov V E, Demokritov S O, Rott K, Krzysteczko P and Reiss G 2007 *Appl. Phys. Lett.* **91** 252504
[17] Clausen P, Vogt K, Schultheiss H, Schäfer S, Obry B, Wolf G, Pirro P, Leven B and Hillebrands B 2011 *Appl. Phys. Lett.* **99** 162505
[18] Vogt K, Schultheiss H, Hermsdoerfer S J, Pirro P, Serga A A and Hillebrands B 2009 *Appl. Phys. Lett.* **95** 182508
[19] Fallarino L, Madami M, Duerr G, Grundler D, Gubbiotti G, Tacchi S and Carlotti G 2013 *IEEE Trans. Magn.* **49** 1033
[20] Kostylev M P, Gubbiotti G, Hu J G, Carlotti G, Ono T and Stamps R L 2007 *Phys. Rev. B* **76** 054422
[21] Madami M, Bonetti S, Consolo G, Tacchi S, Carlotti G, Gubbiotti G, Mancoff F B, Yar M A and Akerman J 2011 *Nature Nanotech.* **6** 635
[22] Kruglyak V V, Keatley P S, Neudert A, Delchini M, Hicken R J, Childress J R and Katine J A 2008 *Phys. Rev. B* **77** 172407
[23] Perzlmaier K, Woltersdorf G and Back C H 2008 *Phys. Rev. B* **77** 054425
[24] Bauer M, Lopusnik R, Dötsch H, Kalinikos B A, Patton C E, Fassbender J and Hillebrands B 2001 *J. Magn. Magn. Mater.* **226** 507
[25] Park J P, Eames P, Engebretson D M, Berezovsky J and Crowell P A 2002 *Phys. Rev. Lett.* **89** 277201

[26] Barman A, Kruglyak V V, Hicken R J, Rowe J M, Kundrotaite A, Scott J and Rahman M 2004 *Phys. Rev. B* **69** 174426
[27] Buttner O *et al* 1998 *IEEE Trans. Magn.* **34** 1381
[28] Damon R W and Eshbach J R 1960 *J. Appl. Phys.* **31** S104
[29] Kwon J, Mukherjee S, Deorani P, Hayashi M and Yang H 2013 *Appl. Phys. A* **111** 369
[30] Sekiguchi K, Yamada K, Seo S M, Lee K J, Chiba D, Kobayashi K and Ono T 2010 *Appl. Phys. Lett.* **97** 022508
[31] Deorani P, Kwon J H and Yang H 2013 *Curr. Appl. Phys.* **14** S129
[32] Bayer C, Park J P, Wang H, Yan M, Campbell C E and Crowell P A 2004 *Phys. Rev. B* **69** 134401
[33] Bayer C, Demokritov S O, Hillebrands B and Slavin A N 2003 *Appl. Phys. Lett.* **82** 607
[34] Bailleul M, Olligs D and Fermon C 2003 *Phys. Rev. Lett.* **91** 137204
[35] Jorzick J, Demokritov S O, Hillebrands B, Bailleul M, Fermon C, Guslienko K Y, Slavin A N, Berkov D V and Gorn N L 2002 *Phys. Rev. Lett.* **88** 047204
[36] Birt D R, An K, Tsoi M, Tamaru S, Ricketts D, Wong K L, Khalili Amiri P, Wang K L and Li X 2012 *Appl. Phys. Lett.* **101** 252409
[37] Donahue M J and Porter D G OOMMF User's Guide, National Institute of Standards and Technology (1999).

**Figure captions**

**Figure 1.** (a) Schematic diagram of the experimental setup. The sample is a Py stripe of dimensions 200 μm × 20 μm. A square electric pulse ($I_{pulse}$) is applied along the 9 μm-wide stripline, and an external bias field $H_b$ along the –x-axis. (b) Time response at a distance of 8 μm from the stripline in the 20 μm-wide Py stripe for a current-generated Oersted field $h_{pulse}$ = 17.9 Oe at $H_b$ = 30 Oe. The open black circles represent the experimental data and the solid red line represents the fit. (c) Frequency spectrum of the time-domain signal revealing two main peaks.

**Figure 2.** (a) Frequency of magnetostatic surface spin wave (MSSW) modes as a function of magnetic field ($H_b$), measured at the center of the stripe. Open circles and squares represent the center mode frequencies, while the solid lines represent the fits according to the MSSW dispersion relation. (b) Variation in the spin wave intensity as a function of distance from the stripline. The solid red line represents the exponential decay fitting to extract the spin wave decay length. The inset shows a typical Gaussian fitting of the measured spin wave packet to extract the spin wave amplitude. (c) Variation in the spin wave intensity as a function of $H_b$. (d) Non-reciprocity factor (κ) as a function of $H_b$. All measurements were performed at $h_{pulse}$ = 17.9 Oe.

**Figure 3.** Spin wave frequencies at different distances from the left edge of the 20 μm wide stripe. The measurements are performed at a distance of 5 μm (a) and 10 μm (b) from the stripline, respectively with a 10 V, 0.2 ns electric pulse.

**Figure 4.** (a) Schematic illustration of the simulated magnetic stripe with a 9 μm-wide linear stripline in the middle of the stripe at $t$ = 3 ns. (b) Simulated profiles of the time-varying magnetization in the y-direction ($M_y$) at different distances from the stripline for $H_b$ = 100 Oe and $h_{pulse}$ = 17.9 Oe. (c) Simulated profile of $M_y$ at a distance of 45.5 μm from the stripline for $H_b$ =

100 Oe. The inset shows frequency spectrum of the simulated time-domain waveform revealing two main peaks corresponding to modes 1 and 2 as indicated.

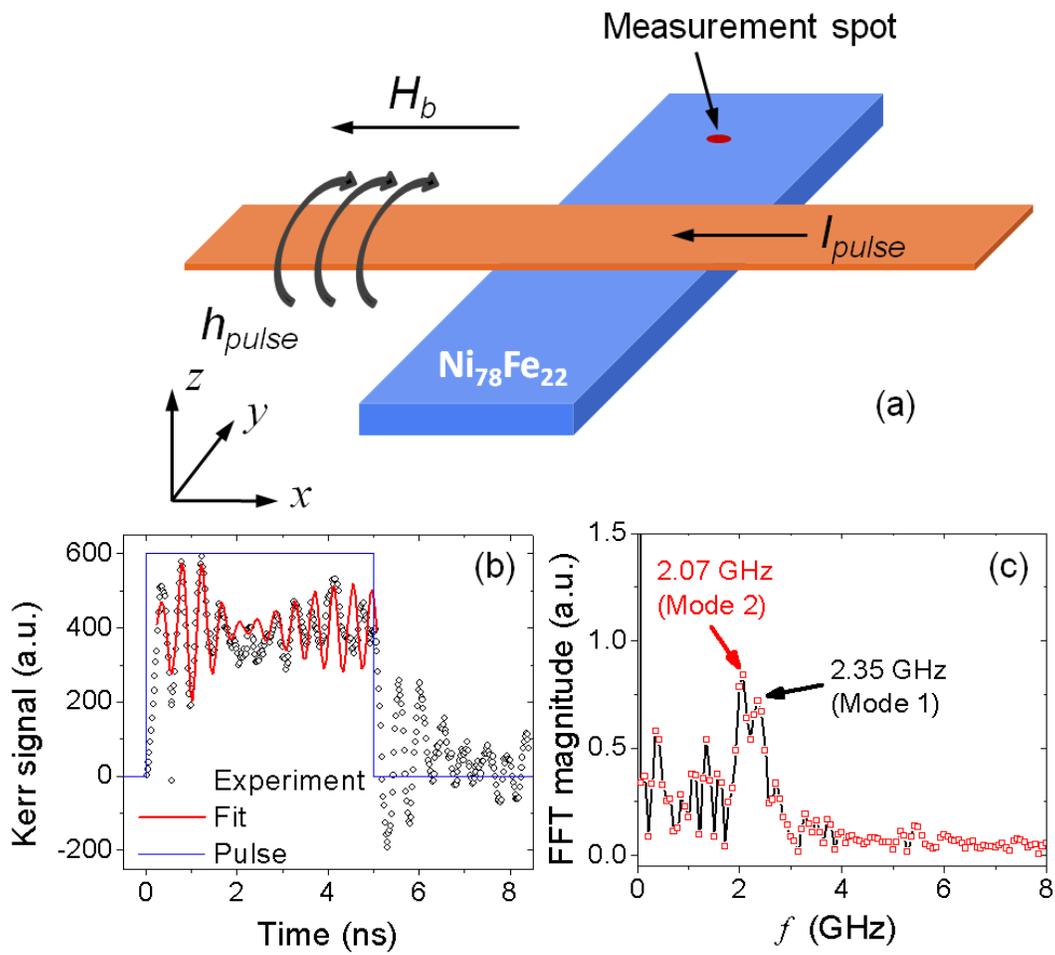

Figure 1

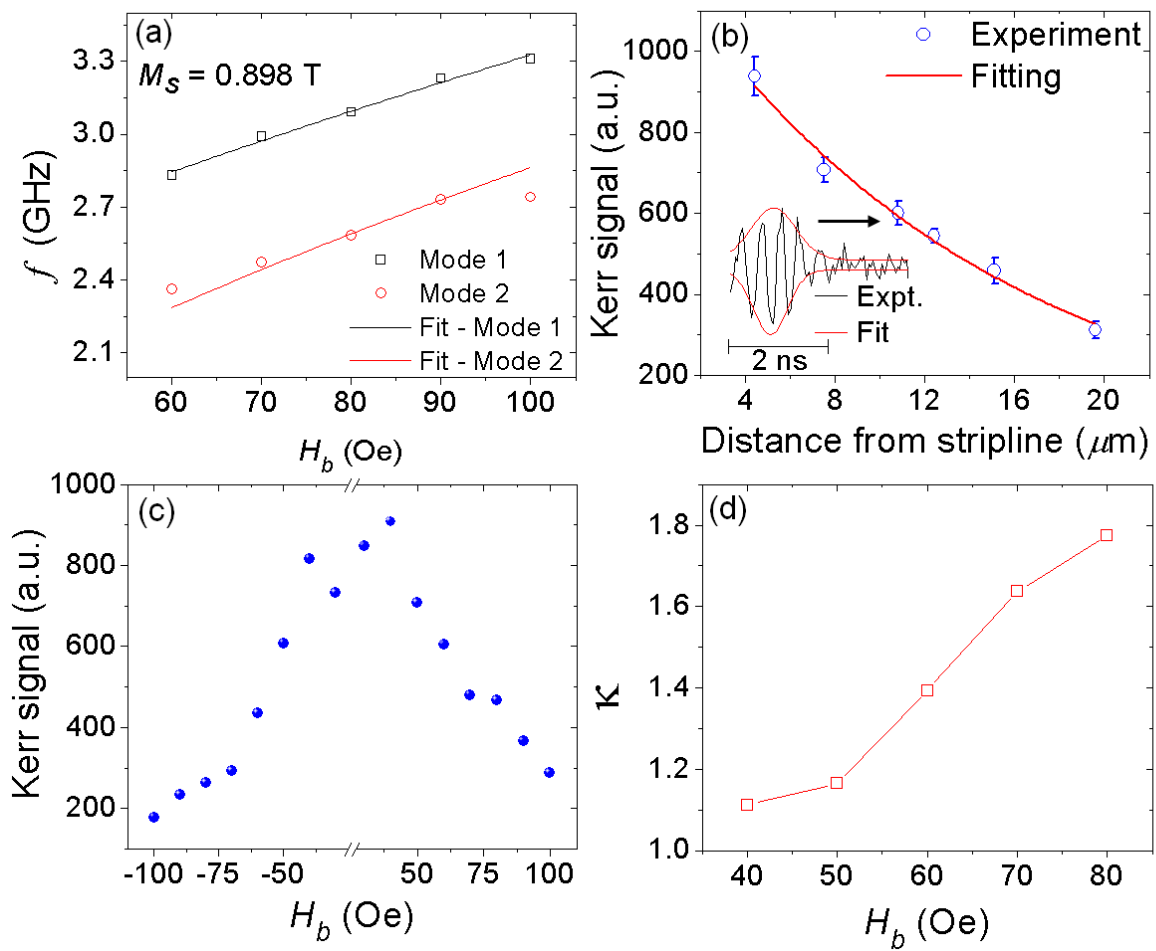

Figure 2

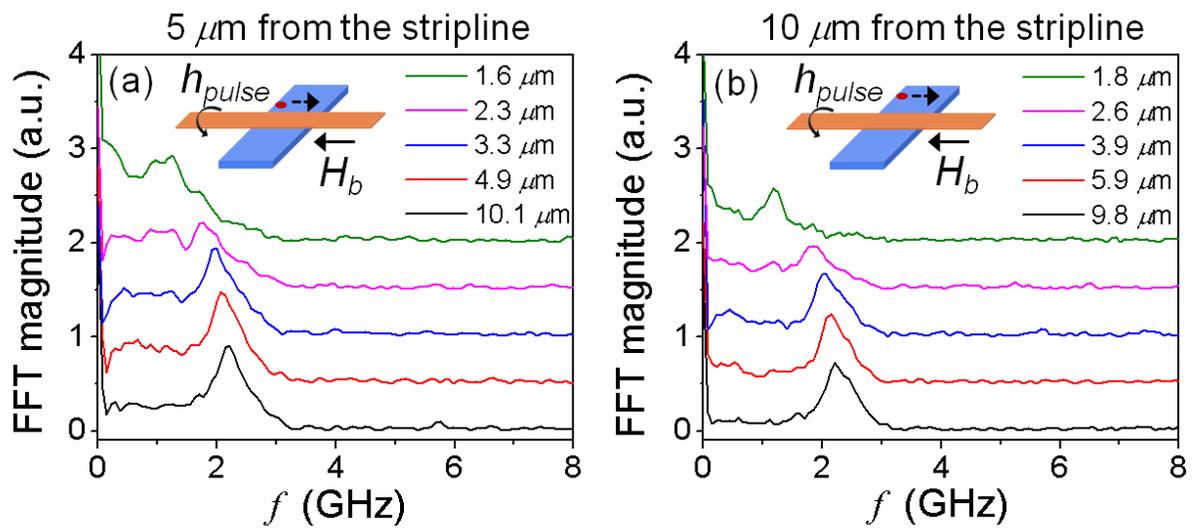

Figure 3

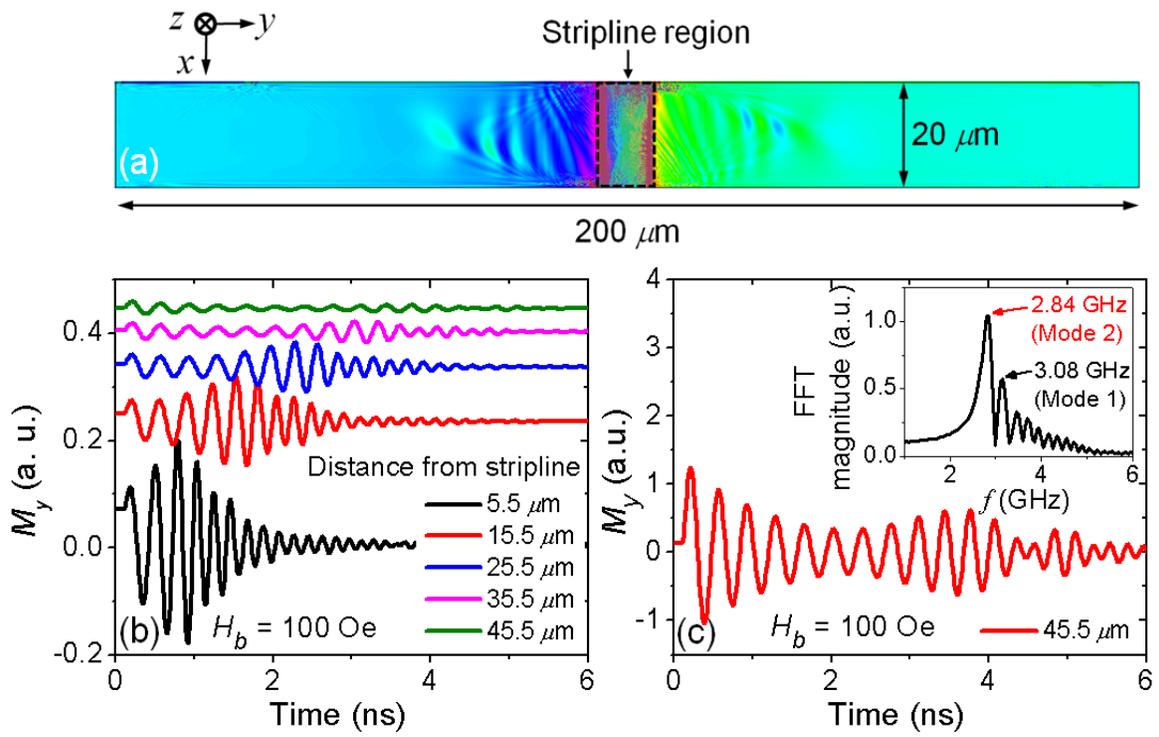

Figure 4